\documentclass[cits]{PoS}
\pdfoutput=1

\usepackage{times}
\usepackage[T1]{fontenc}
\usepackage{graphicx}
\usepackage{amsmath}

\newcommand{\Tr}{\mathrm{Tr}}

\newcommand{\MeV}{~\mathrm{MeV}}

\newcommand{\fm}{~\mathrm{fm}}

\newcommand{\SU}{\mathrm{SU}}

\newcommand{\D}{\mathrm{D}}

\newcommand{\lk}{\left<}

\newcommand{\rk}{\right>}

\newcommand{\appropto}{\mathrel{\vcenter{
  \offinterlineskip\halign{\hfil$##$\cr
    \propto\cr\noalign{\kern2pt}\sim\cr\noalign{\kern-2pt}}}}}

\title{Lattice QCD with 2+1 Flavors and Open Boundaries: First Results of the Baryon Spectrum}

\ShortTitle{Lattice QCD with 2+1f and Open Boundaries: First Results of the Baryon Spectrum}

\author{\speaker{Wolfgang S\"oldner}{ [for the RQCD collaboration]}\thanks{
We acknowledge PRACE for awarding us access to FERMI based in Italy at
CINECA, Bologna and to SuperMUC based in Germany at LRZ, Munich, as well as to JUQUEEN at J\"ulich
Supercomputing Centre (JSC).
Support by the DFG SFB/TRR-55 ``Hadron Physics from Lattice QCD'' is gratefully acknowledged.
}\\
        Institut f\"ur Theoretische Physik, Universit\"at Regensburg, \\D-93040 Regensburg, Germany
\\
        E-mail: \email{wolfgang.soeldner@physik.uni-regensburg.de}
        }

\abstract{Based on CLS simulations with 2+1 flavors and open boundaries we present first results of the baryon spectrum. We report on the status of our effort related to these simulations and the chiral extrapolation to the physical point.}

\FullConference{The 32nd International Symposium on Lattice Field Theory,\\
		23-28 June, 2014\\
		Columbia University New York, NY}

\begin{document}

\section{Introduction}
As more computing power and better algorithms have become available over the years the precision in lattice QCD results becomes more and more relevant.
In order to have small errors on lattice QCD results it is important to have good control over the systematics involved in lattice simulations.
Besides large volumes, quark masses close to the physical point, and large statistics, the continuum limit plays a prominent role because of the so-called {\it topological freezing problem}~\cite{DelDebbio:2002xa, Bernard:2003gq, Schaefer:2010hu}.
As the lattice spacing is decreased and, hence, the continuum is approached it becomes more and more difficult for conventional simulation techniques to tunnel between topologically distinct gauge sectors
during the simulation. As a result, one faces large autocorrelation times for simulations at lattice spacings smaller than about $0.05\fm$~\cite{Bruno:2014ova}.
With the conventional approach large statistics and/or large spatial volumes would be necessary to obtain meaningful results. A quite different and new approach to reduce the large autocorrelation times for
quantities which couple to those topological modes is based on the introduction of open boundary conditions~\cite{Luscher:2011kk} which are, in practice, imposed in the time direction. Topological objects can then flow in and out through the temporal boundary,
and, in fact, it has turned out that this approach is computationally advantageous~\cite{Luscher:2012av} with respect to an -- otherwise necessary -- increase in volume and/or statistics.

Therefore, the CLS collaboration~\cite{CLS} has started a major effort to generate gauge field configurations with open boundary conditions aiming at large statistics and volumes as well as pion masses close to the physical point in
order to get good control over all systematics. There has been generated a relatively large set of ensembles quite recently using Wilson fermions with $2+1$ flavors~\cite{Bruno:2014jqa}.
We report on first measurements on some of these ensembles,  in particular we focus on first results of the $\mathrm{SU}(3)$ octet baryon spectrum and also briefly address the scale setting,
see Sec.~\ref{sec:baryon} and Sec.~\ref{sec:scale}, respectively. Before, we will start with a short presentation of some details about the simulations.

\section{Simulation Details\label{sec:detail}}
  \begin{table}[bf]
  \centering
  \begin{tabular}{c|c|c|c|c|c}
  run id & H101 & H102 & H105 & C101 &  D100 \\
  \hline
  $m_\pi \approx$  & 420 MeV & 350 MeV & 280 MeV & 220 MeV & 130 MeV \\
  \hline
  $m_K\approx$    & 420 MeV & 440 MeV & 460 MeV & 470 MeV & 480 MeV
  \end{tabular}
  \caption{\label{tab:masses}CLS ensembles at $\beta=3.4$. Note that the D100 ensemble has currently only low statistics and is not considered in the analysis presented in this contribution.}
  \end{table}
A non-perturbatively $\mathcal{O}(a)$-improved Wilson action with tree-level Symanzik improved gauge action and $2+1$ flavors has been employed.
We have simulated at pion masses ranging from about $420\MeV$ down to about $130\MeV$ following the strategy adopted from the QCDSF collaboration~\cite{Bietenholz:2011qq} where the sum of bare quark masses is kept constant
along the chiral trajectory. This strategy not only simplifies the tuning process of the simulations but also has the advantage that one can extrapolate,~e.g.,~the combination $m_K^2+m_\pi^2/2$ rather precisely to the physical point as
it turns out that in leading order chiral perturbation theory this combination is proportional to the sum of quark masses,  $m_K^2+m_\pi^2/2 \propto \Tr M \equiv \sum_i m_i$, and, in addition, the corrections to this relation
stay small down to the physical point. As the scale $t_0$ is known to depend only weakly on the quark mass (see also the upper left plot in Fig.~\ref{fig:chiral})
we use the dimensionless combination $\phi_4 = 8 t_0 (m_K^2 + m_\pi^2/2)$ and match our chiral trajectories with different lattice spacings at a single point along the symmetric line, i.e.~we have fixed $\Tr M$, respectively $\sum_i 1/\kappa_i$, 
for each lattice spacing to match our target estimate $\phi_4 |_{m_{ud}=m_s}=1.15$ for $m_{ud} \equiv m_u =m_d = m_s$. Note that, at the physical point, we obtain $\phi_4^{\mathrm phys} |_{m_{ud}=m_s}=1.117(38)$ with 
$m_\pi  = 134.8(3) \MeV$ and $m_K = 494.2(4) \MeV$ from Ref.~\cite{Aoki:2013ldr}, and $\sqrt{8t_0} = 0.4144(59)(37) \fm$ from Ref.~\cite{Borsanyi:2012zs}.

    \begin{figure}[tf]
    \centering
      \includegraphics[width = 0.49\textwidth]{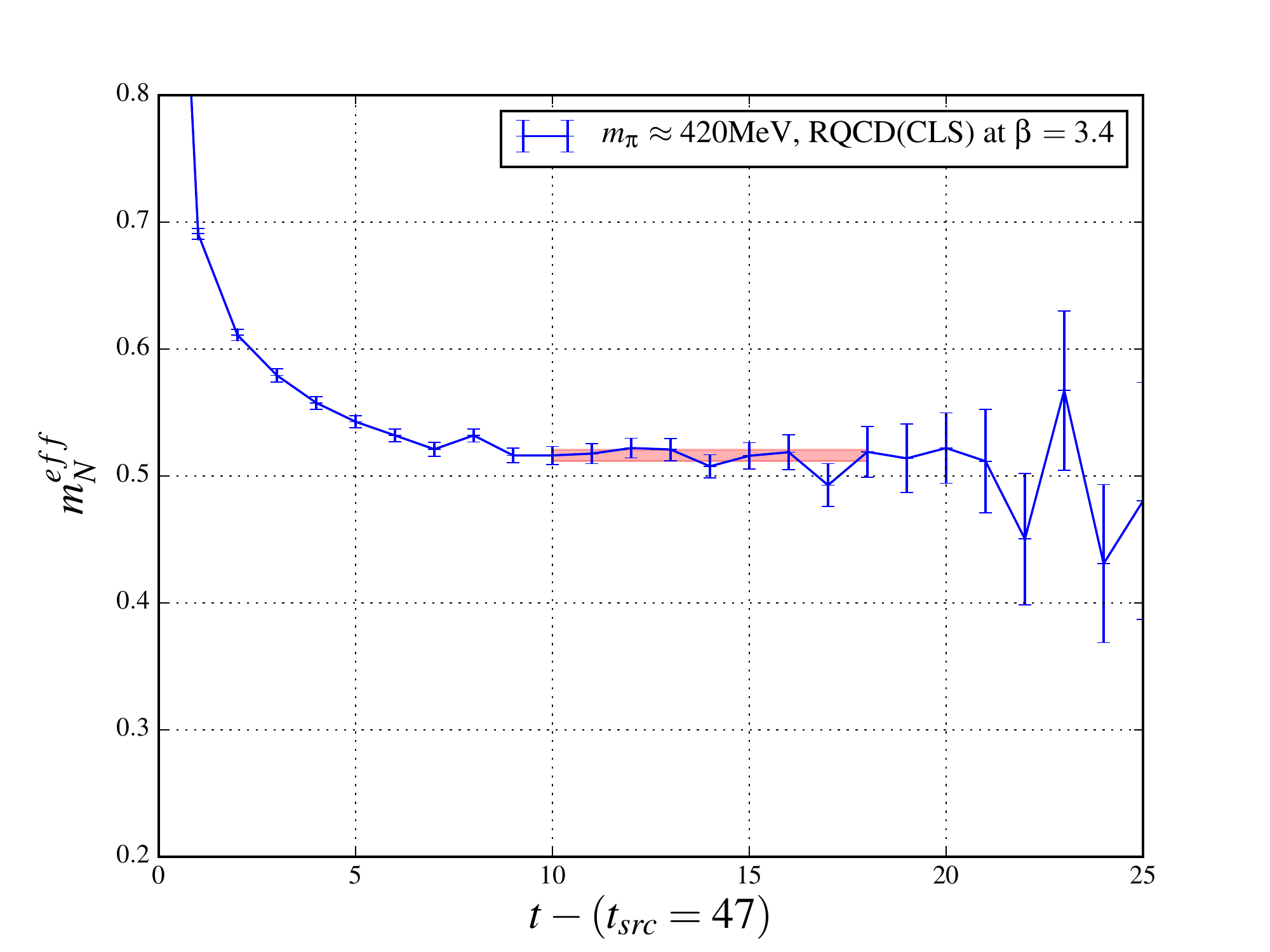}
      \includegraphics[width = 0.49\textwidth]{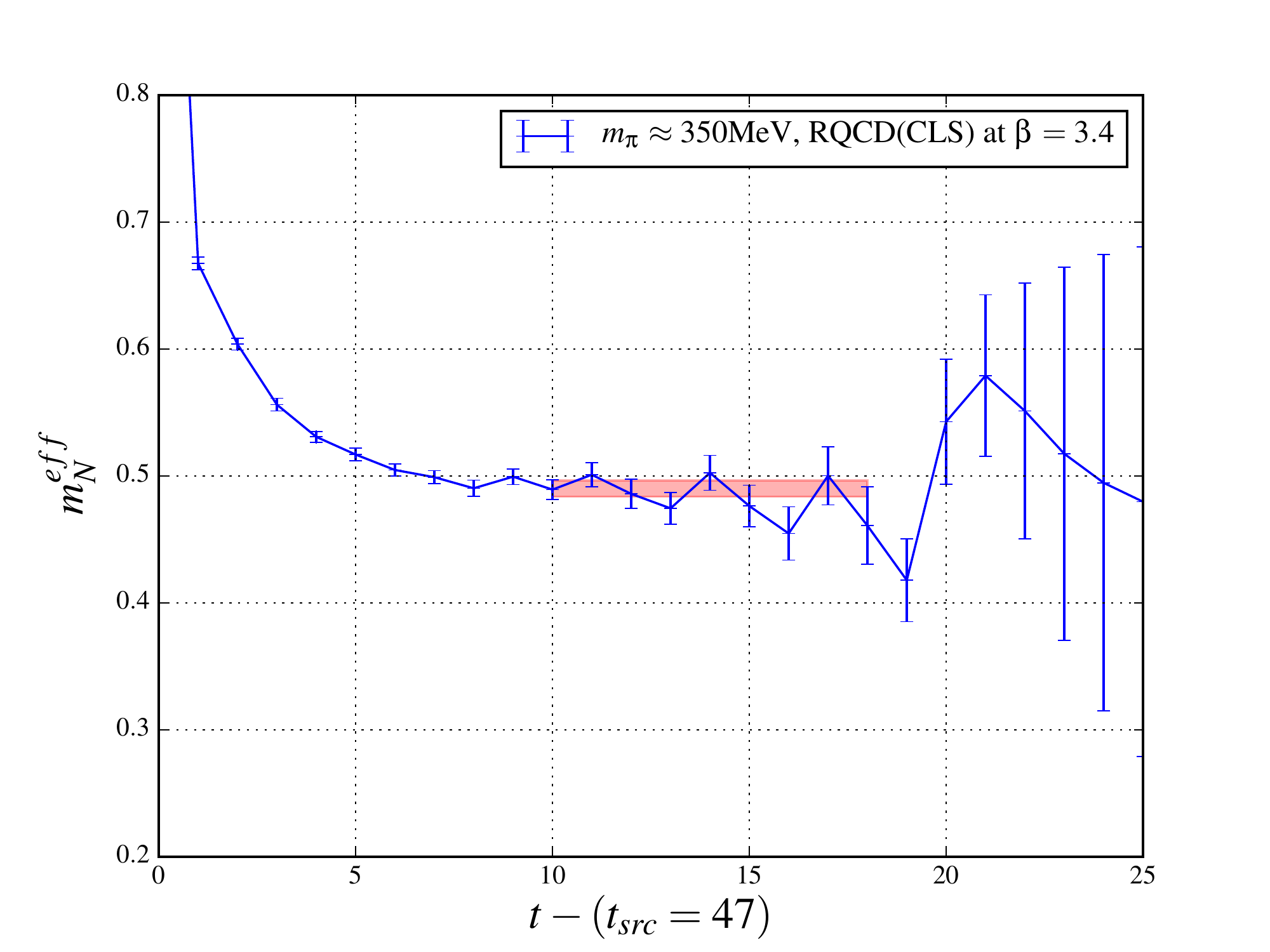}\\
      \includegraphics[width = 0.49\textwidth]{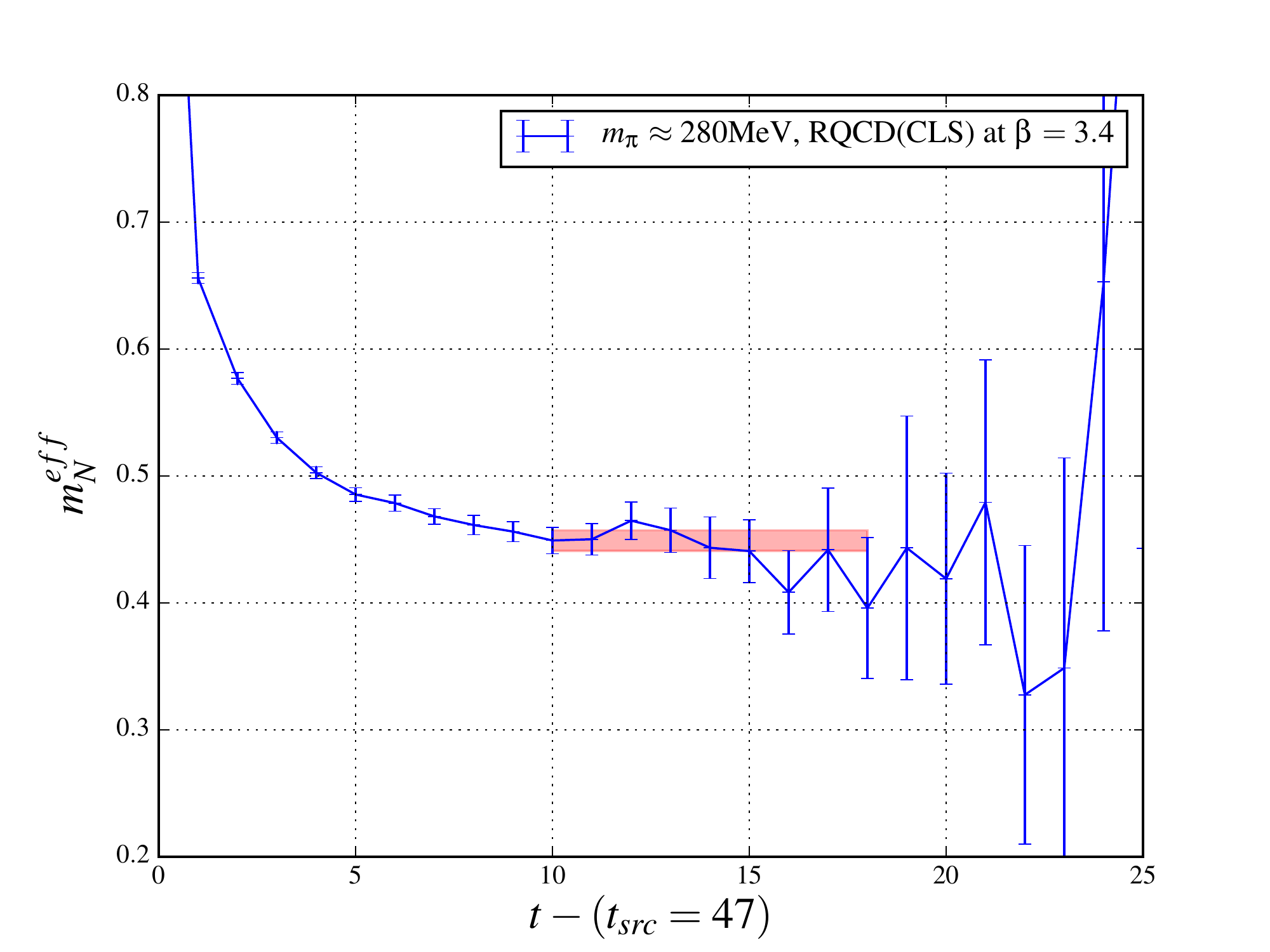}
      \includegraphics[width = 0.49\textwidth]{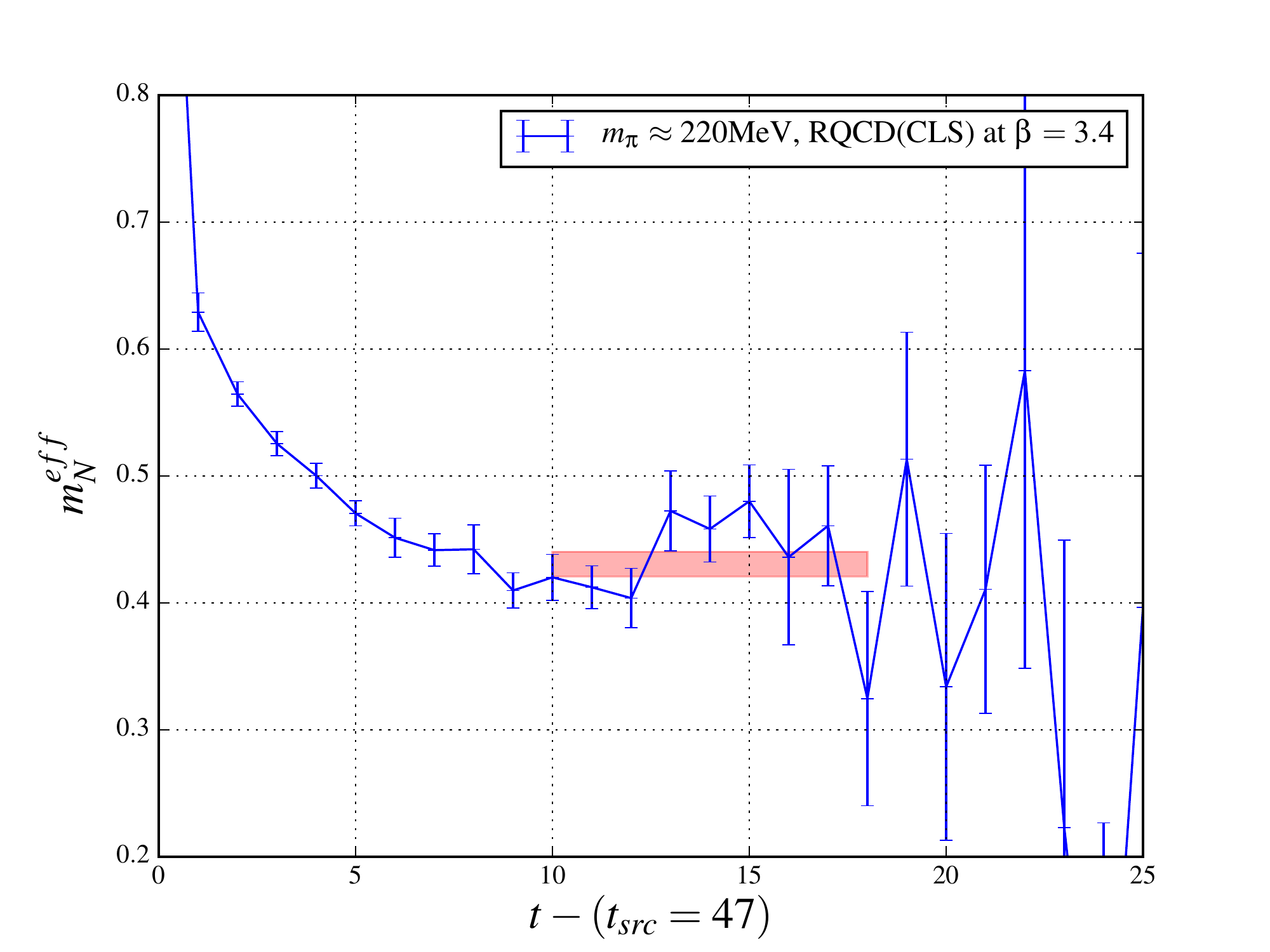}
       \caption{\label{fig:baryon}Effective mass plots of the nucleon.
        The horizontal length of the red bars in the plots indicate the fit range which was chosen to be~$[10,18]$, the vertical position indicates the fitted mean value and the width corresponds to the fit error.
        Note that for the C101 ensemble ($m_\pi\approx 220 \MeV$) the statistics is smaller compared to the other ensembles.}
    \end{figure}

Currently, we have generated ensembles at three different lattice gauge couplings with $\beta = 6/g^2 = 3.4, 3.55, 3.7$. In the following we present preliminary data only for our ensembles at $\beta = 3.4$ which corresponds to
a lattice spacing $a \approx 0.086 \fm$. Note that also for $\beta = 3.55, 3.7$ other quantities have been presented at this conference~\cite{Bruno:2014lra,Korcyl:2014tla}.     
We have summarized the pion and kaon masses of the ensembles with $\beta=3.4$ in Tab.~\ref{tab:masses}. Note that the D100 ensemble has currently only low statistics and is therefore not be taken into account in the analysis below.
The lattice geometry for the ensembles H101, H102, and H105 is $32^3 \times 96$ and for the C101 ensemble $48^3 \times 96$ where the temporal extent of the lattices is typically larger compared to simulations with periodic boundary conditions
due to the boundary effects specific to open boundary conditions. All ensembles have large spatial volumes $V_s=L^3$; for all ensembles $m_\pi L \gtrsim4$ is satisfied. The HMC trajectory has length $\tau=2$ and for the ensembles under
consideration more than 4000 Molecular Dynamic Units (MDUs) for each ensemble have been generated.

For the generation of gauge field configurations the openQCD package~\cite{openQCD} has been used which has several algorithmic improvements built in. Besides the Hasenbusch trick~\cite{Hasenbusch:2001ne}, improved integrators~\cite{Omelyan2003272}, a multi-level integration scheme~\cite{Sexton:1992nu},
and a deflated solver~\cite{Luscher:2007es,Frommer:2013fsa}, the twisted mass reweighting~\cite{Luscher:2012av} should be pointed out because it is the first time that this technique has been applied within a larger set of simulations. In the light fermion part of the action a twisted mass
term is introduced in order to push the eigenvalues of the Dirac operator away from zero and, hence, increase the stability of the HMC simulation. 

To correct for the additional twisted mass term in the light quark action one has to reweight the observables. In addition, one also applies reweighting with respect to the strange quark action
to account for errors coming from the rational approximation which is used to simulate the strange quark. The ensemble average of an observable
	  $\lk O \rk = \frac{\lk W O \rk}{\lk W \rk}$ then includes the reweighting factor $W=W_0 W_1$ with a light and strange reweighting factor $W_0$ and $W_1$, respectively.
 We found that these reweighting methods work quite nicely in practice, more details can be found in Ref.~\cite{Bruno:2014jqa}.

\section{Baryon Spectrum\label{sec:baryon}}
 
In this section we present first preliminary results on the baryon spectrum from simulations with open boundary conditions. In Sec.~\ref{sec:baryon_comp} we give some details about the computation of the spectrum, in Sec.~\ref{sec:baryon_chiral}
we discuss the chiral extrapolation of our results and also compare our findings to results from the literature.

    \begin{figure}[t]
      \includegraphics[width = 0.5\textwidth]{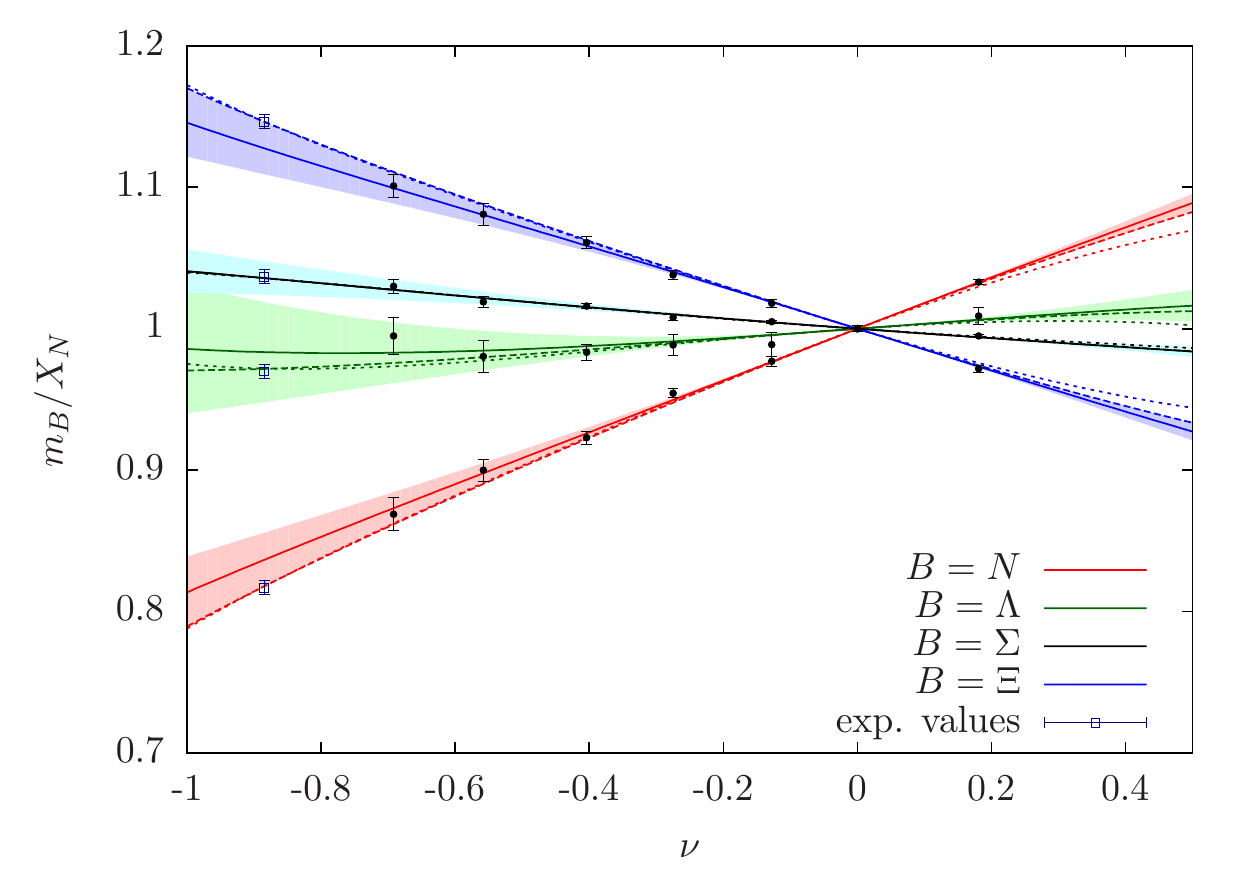}
      \includegraphics[width = 0.5\textwidth]{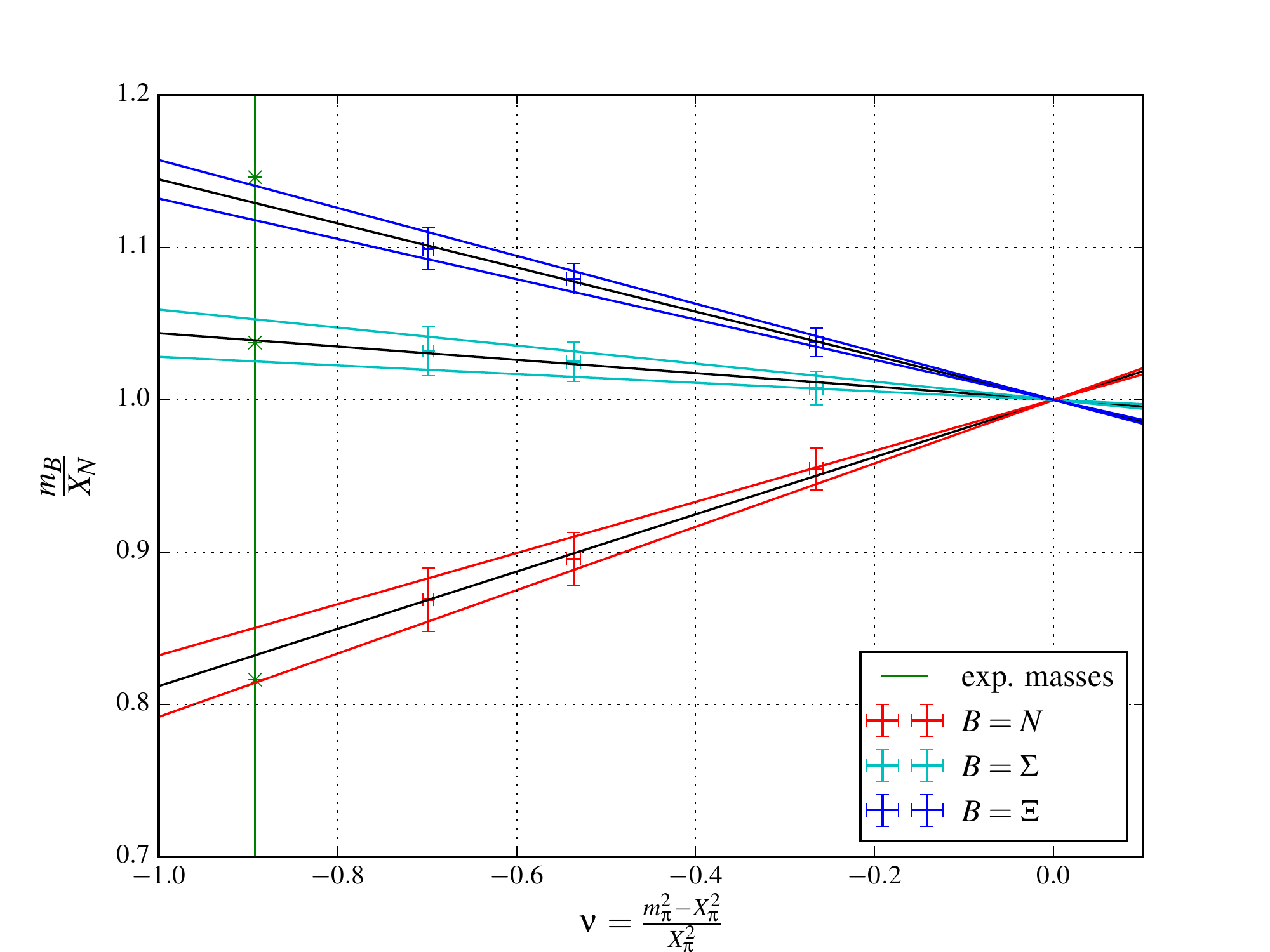}
      \caption{\label{fig:fanplot}Fits to the octet baryon masses of our preliminary data (right) compared to fits from Ref~\cite{Bietenholz:2011qq} based on data from Ref.~\cite{Bruns:2012eh} (left plot was taken from Ref~\cite{Bietenholz:2011qq}).}
    \end{figure}

\subsection{Computational Details\label{sec:baryon_comp}}

For the computation of the baryon spectrum we have used standard relativistic interpolators, e.g.~for the nucleon we have used $I_N = \epsilon_{abc} u_a \left( u_b^T C \gamma_5 d_c\right)$. The source position in temporal direction was always
chosen to be at the center of the lattice, i.e.~$t_{\mathrm src}=47$ where $t \in [0,\dots,95]$, the spatial source position was chosen randomly. We have been investigating smeared-smeared point-to-all correlators where we have used
100 steps of Wuppertal smearing on APE smeared gauge links for both the source and the sink. The fit range was chosen to be $t \in [10,18]$. For the ensembles H101, H102, and H105 all available 4000 MDUs for each ensemble have been analyzed,
for the C101 ensemble 1000 MDUs have been considered so far.
For the analysis a custom version of the CHROMA software package~\cite{Edwards:2004sx} has been developed where also the deflated solver~\cite{Luscher:2007es} is used.

In Fig.~\ref{fig:baryon} we show the effective mass of the nucleon for the ensembles under consideration. For the three ensembles where all configurations have been analyzed we observe a nice plateau while for the C101 ensemble one
would expect that a less noisy plateau will develop once full statistics have become available. For $\Sigma$ and $\Xi$ the corresponding plots look similar.
Note that it has turned out that for this preliminary analysis the smearing was not chosen optimally. We are currently running measurements with full statistics and optimized smearing and, hence, we expect to improve on the errors present in this study.

\subsection{Chiral Extrapolation\label{sec:baryon_chiral}}
In order to chirally extrapolate to the physical point we make use of $\SU(3)$ chiral perturbation theory. The octet baryon masses in terms of the pion and kaon masses to order $\mathcal{O}(p^2)$ are given by~\cite{Bernard:1993nj},
\begin{align}
\begin{split}
m_{N} &= m_{0}-4 b_{D} \dot{M}_{K}^2+4 b_{F} \left(\dot{M}_{K}^2-\dot{M}_{\pi}^2\right)-2 b_{0} \left(2 \dot{M}_{K}^2+\dot{M}_{\pi}^2\right)+\cdots\,,\\   
m_{\Lambda} &= m_{0}+\frac{4}{3} b_{D} \left(-4 \dot{M}_{K}^2+\dot{M}_{\pi}^2\right)-2 b_{0} \left(2 \dot{M}_{K}^2+\dot{M}_{\pi}^2\right)+\cdots\,, \\
m_{\Sigma} &= m_{0}-4 b_{D} \dot{M}_{\pi}^2-2 b_{0} \left(2 \dot{M}_{K}^2+\dot{M}_{\pi}^2\right)+\cdots\,, \\
m_{\Xi} &= m_{0}-4 b_{D} \dot{M}_{K}^2-4 b_{F} \left(\dot{M}_{K}^2-\dot{M}_{\pi}^2\right)-2 b_{0} \left(2 \dot{M}_{K}^2+\dot{M}_{\pi}^2\right) +\cdots\,.
\label{eq:baryon_chiral}
\end{split}
\end{align}
The average nucleon mass $X_N$ and average pion mass $X_\pi$ are defined by $X_N \equiv \left( m_N + m_\Sigma + m_\Xi  \right)/3$ and $X_\pi^2 \equiv \left( 2 m_K^2 + m_\pi^2  \right)/3$. In leading order chiral perturbation theory one finds,
\begin{equation}
X_ N = m_{0}-2 b_{0} \left(2 \dot{M}_{K}^2+\dot{M}_{\pi}^2\right)+\cdots \,.
\end{equation}
Since $2 m_{K}^2+m_{\pi}^2$ is approximately constant along our chiral trajectory also the average nucleon mass stays constant to leading order along that line. We define the dimensionless quantity $\nu=\tfrac{m_\pi^2-X_\pi^2}{X_\pi^2}$ and plot the
fits of our data to Eq.~\eqref{eq:baryon_chiral} in the right plot of Fig.~\ref{fig:fanplot}. The errors on the masses come from uncorrelated fits where we have performed a jackknife analysis with binning.
The left plot of Fig.~\ref{fig:fanplot} is taken from~\cite{Bruns:2012eh} where chiral fits have been performed to data from QCDSF~\cite{Bietenholz:2011qq}.
Comparing those to our preliminary results we find nice agreement.

\section{Scale Setting\label{sec:scale}}
To set the scale at a certain lattice coupling one would choose a dimensionful, (ideally) physical quantity and compare it to the lattice result. In order to do so we have to, in our case, chirally extrapolate to the physical point. 
For the following discussion we define as the physical point the value of $\kappa_l \equiv \kappa_l^{\mathrm phys}$ 
at which the ratio of strange to light $\mathcal{O}(a)$-improved AWI quark masses takes its physical value of $27.46(44)$~\cite{Aoki:2013ldr}. The preliminary value of $\kappa_l^{\mathrm phys}$ is taken from Ref.~\cite{ws}.
Note that $\kappa_s^{\mathrm phys}$ is then determined by our definition of the chiral trajectory through $\Tr M = \mathrm{const}$. We remark that this definition is preliminary
since we would hope to improve over the errors on that ratio given in Ref.~\cite{Aoki:2013ldr} once the continuum limit can be taken.
	\begin{figure}[t]
	\centering
	  \includegraphics[width = .49\textwidth]{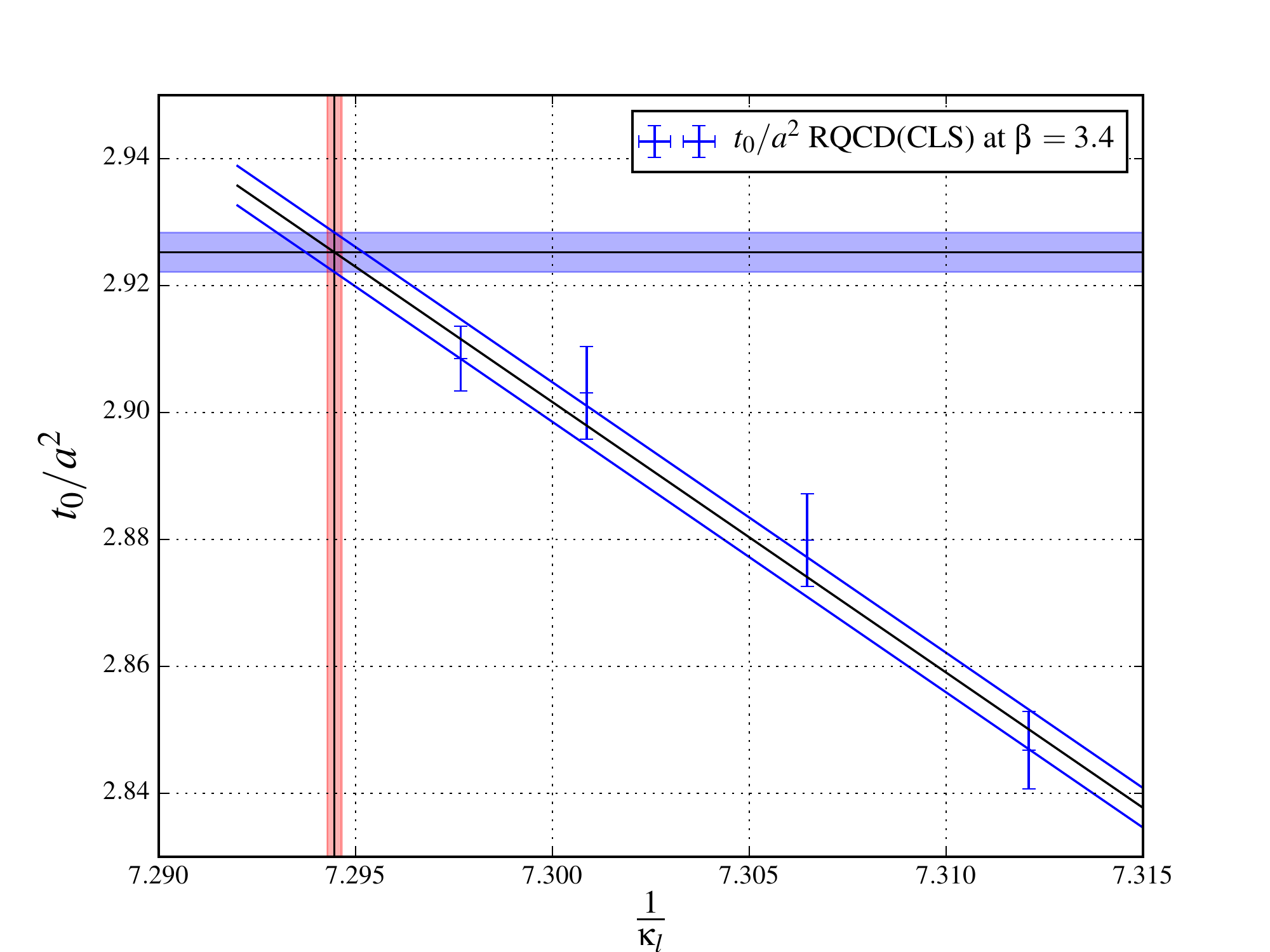}
	  \includegraphics[width = .49\textwidth]{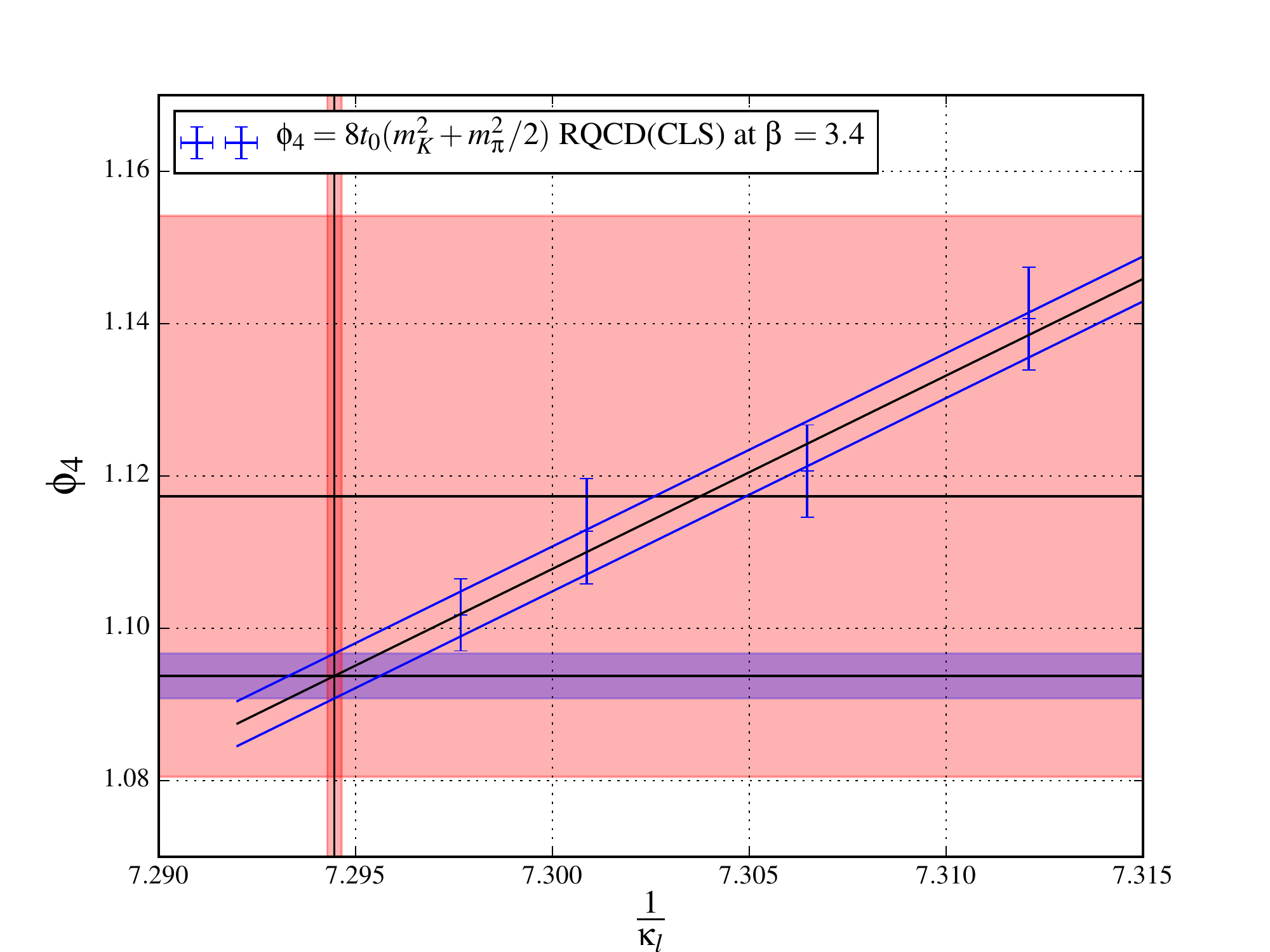} \\
          \includegraphics[width = .49\textwidth]{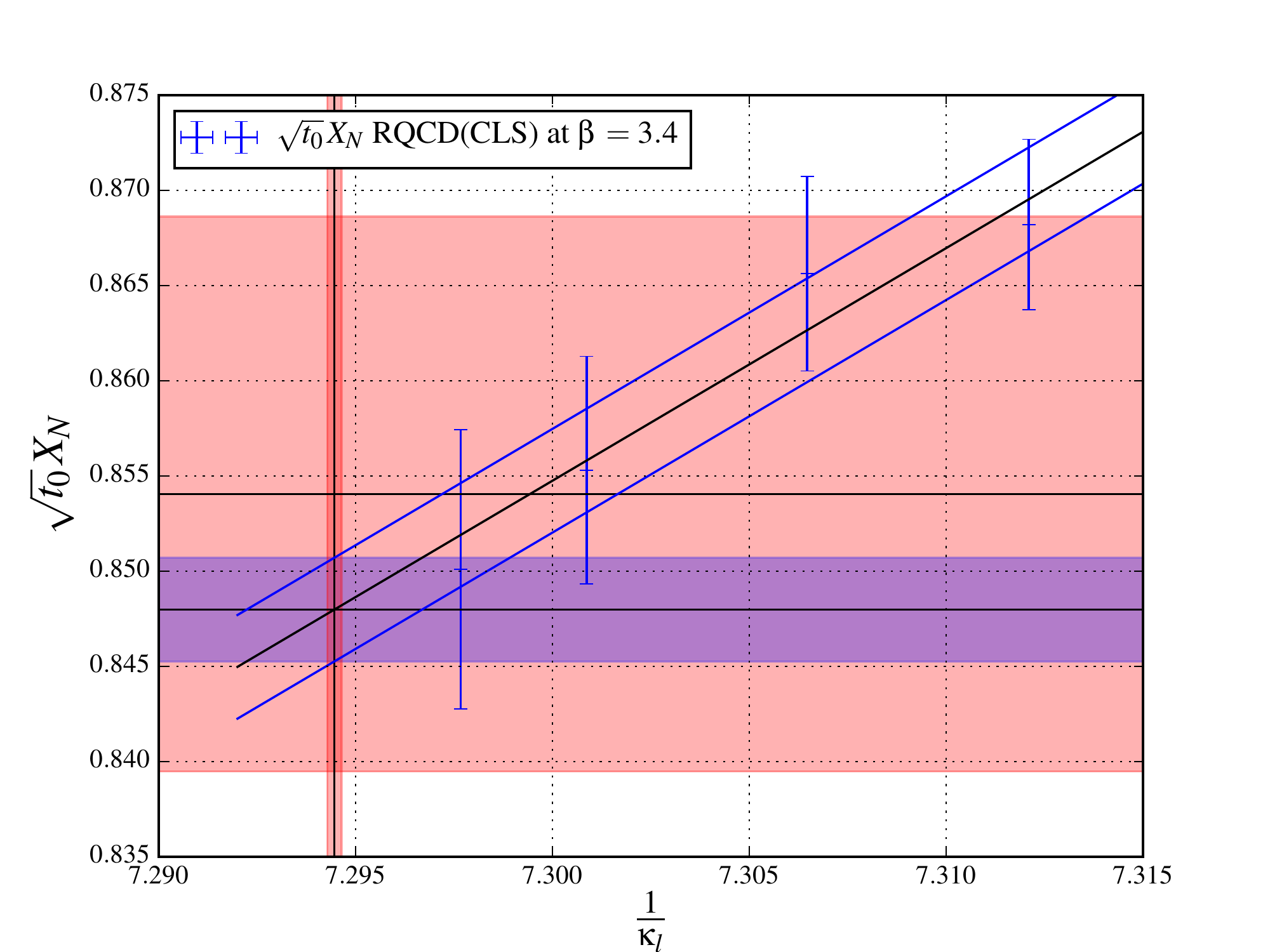} 
          \includegraphics[width = .49\textwidth]{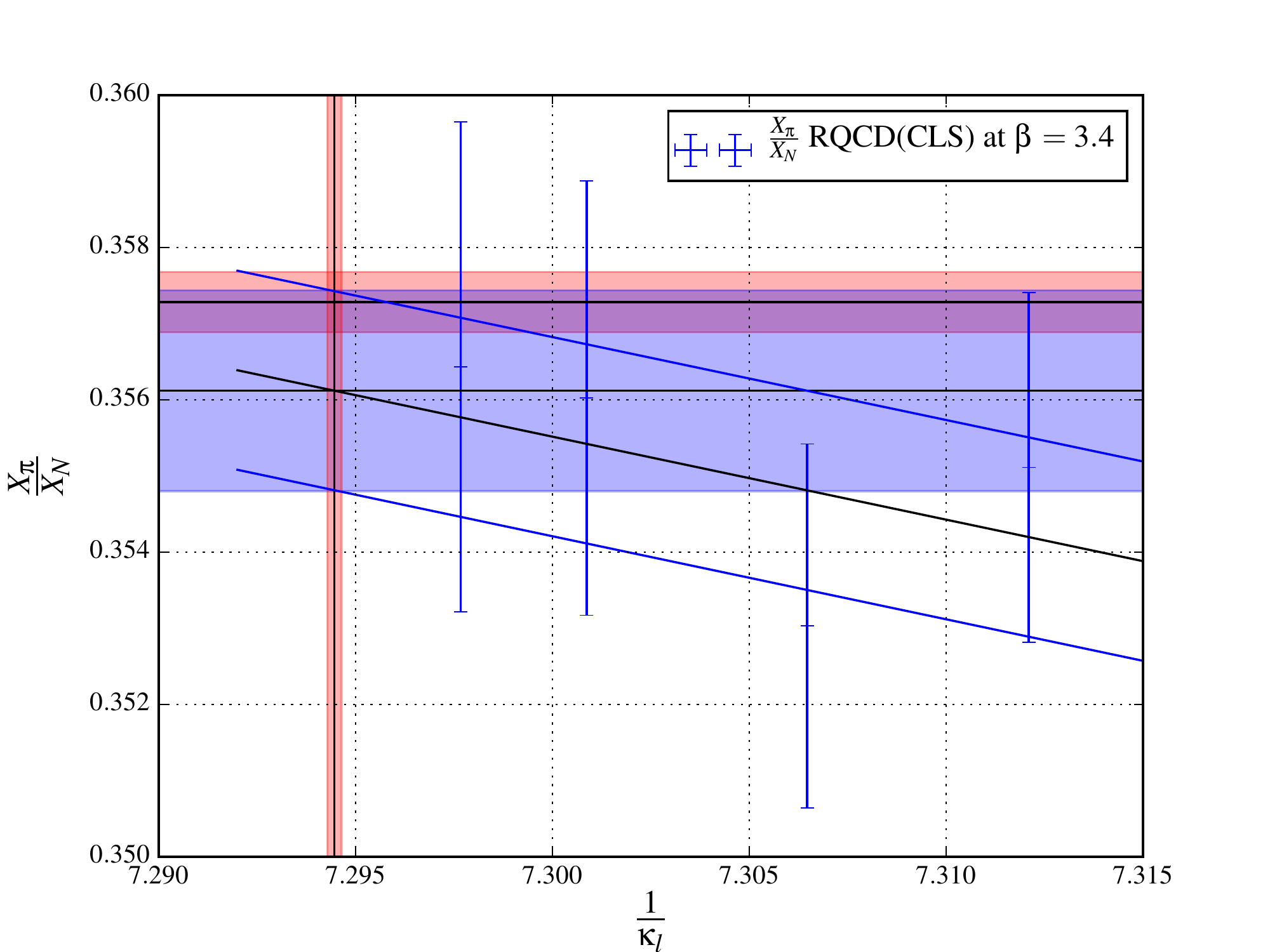}
	  \caption{\label{fig:chiral}Linear extrapolations of $\tfrac{t_0}{a^2}$, $\phi_4$, $\sqrt{t_0} X_N$, $\tfrac{X_\pi}{X_N}$ to the physical point (vertical red bars). The horizontal red bars indicate
	  the corresponding experimental value including errors (or respectively the theoretical value including errors in the case of combinations involving $t_0$),
	  the blue horizontal bar represent the respective lattice value with errors obtained by extrapolating the data to the physical point.
	  }
	\end{figure} 

We first look at the chiral extrapolation of $\phi_4$, see the upper right plot in Fig.~\ref{fig:chiral}. Remember that $\phi_4$ has been used to match our chiral trajectory along the symmetric line. We can now ask whether our matching condition
at that point results in an agreement at the physical point, i.e.~$\phi_4$ obtains its physical value at the physical point which is actually what is desired.
Since $m_K^2+m_\pi^2/2$ is constant to leading order along that trajectory and in addition $t_0$ is also only weakly dependent
on the quark mass (see upper left plot in Fig.~\ref{fig:chiral}) we do not expect a strong quark mass dependence of $\phi_4$, in fact this is evident from the upper right plot of Fig.~\ref{fig:chiral} and we find agreement of $\phi_4$ at the 
 physical point with its physical value within errors. Up to cutoff effects, this means that we have indeed chosen the correct value of $\Tr M$ for our chiral trajectory. However, since we do not know much about the cutoff dependence at the present stage, 
it is interesting to compare how different quantities match at the physical point because different quantities could have very different cutoff dependence. Let us take a closer look at $\sqrt{t_0} X_N$ and $\tfrac{X_\pi}{X_N}$.
The chiral extrapolations are shown in the lower plots of Fig.~\ref{fig:chiral}. Within errors we find again agreement for those quantities at the physical point with the corresponding physical values.
Again we would conclude that we have chosen the correct value of $\Tr M$ for our chiral trajectory, and, in addition, the cutoff effects seem to be small.
Note, however, that without knowing the explicit cutoff dependence of the quantities under investigation it is not really possible to disentangle cutoff effects and effects stemming from a mistuning of our chiral trajectory, i.e.~a wrong value of $\Tr M$.
This will be clarified by a future study of our ensembles at different lattice spacings which will also enable an independent scale setting.

We remark that the conclusions we have drawn above, of course, depend on the actual value and error on $t_0$ which is a result from a lattice study~\cite{Borsanyi:2012zs}
as $t_0$ is not experimentally accessible. We therefore also plan to extract $t_0$ from our simulations and once the continuum limit can be performed we expect to deliver an independent cross-check on that quantity.

Finally, coming back to the issue of scale setting, we remark that all quantities discussed in this section can be used for scale setting. As we have found that all these quantities agree well at the physical point with their 
corresponding physical value it is evident that the lattice spacings extracted by the quantities investigated here will results in similar values. As our results are still preliminary we will refrain from quoting numbers but will return
to this topic in a different publication.

\section{Summary and Outlook}
We have presented first results on the baryon octet spectrum from lattice simulations with $2+1$ flavors and open boundaries which has been generated within the CLS collaboration. Using leading order chiral perturbation theory
we have chirally extrapolated results down to the physical point for $N, \Sigma, \Xi$ at pion masses ranging from $m_\pi \approx 220-420 \MeV$ at a lattice spacing $a \approx 0.086 \fm$. We have found nice agreement with the literature. 
In addition, we have extrapolated $\tfrac{t_0}{a^2}$, $\phi_4$, $\sqrt{t_0} X_N$, $\tfrac{X_\pi}{X_n}$ to the physical point and have checked that $\phi_4$, which is used to match our chiral trajectories, agrees at this point with its physical value.
Furthermore, we have found that $\sqrt{t_0} X_N$ and $\tfrac{X_\pi}{X_n}$ also agree with their corresponding physical values at this point suggesting that cutoff effects may be small and that the chiral trajectory has been chosen appropriate.
A more detailed analysis including cutoff effects will follow in the future.

\bibliography{lat14_v2}{}
\bibliographystyle{JHEP}

\end{document}